\documentstyle{article}
\newpage
\begin{document}
%%%%%%%%%%%%%%
% used symbols
%%%%%%%%%%%%%

\def\msolar{{M_\odot}}
\def\lsolar{{L_\odot}}
\def\oversim#1#2{\lower0.5pt\vbox{\baselineskip0pt \lineskip-0.5pt
     \ialign{$\mathsurround0pt #1\hfil##\hfil$\crcr#2\crcr\sim\crcr}}}
\def\gsim{\mathrel{\mathpalette\oversim>}}    % > over \sim
\def\lsim{\mathrel{\mathpalette\oversim<}}    % < over \sim

%%%%%%%%%%%%%%%%%%%%%%%%%%%%%%%%%%%%%%%%%%%%%%%%%%%%%%%%%%%%%%%%%%%%%%
%
%             begin paper
%
%%%%%%%%%%%%%%%%%%%%%%%%%%%%%%%%%%%%%%%%%%%%%%%%%%%%%%%%%%%%%%%%%%%%%%

\begin {center}
{\Large \bf  Stellar evolution on the Asymptotic Giant Branch}
\end {center}
\vskip0.5truecm
\begin {center}
Albert A. Zijlstra
\end {center}
\vskip0.5truecm
\begin {center}
European Southern Observatory, Karl Schwarzschild Strasse
2,\\
D-85748 Garching bei M\"unchen, Germany\\
\vskip0.1truecm
email: {\tt azijlstr@eso.org}
\end {center}
\vskip0.5truecm
\begin {center}
{\it Review paper given at Edinburgh, August 1994, ``Circumstellar
Matter''}
\end {center}
\vskip2.0truecm
\begin{abstract}
Mass loss dominates the stellar evolution on the Asymptotic Giant
Branch. The phase of highest mass-loss occurs during the last 1--10\%
of the AGB and includes the so-called Miras and OH/IR stars. In this
review I will discuss the characteristics and evolution of especially
Miras, and discuss how they are linked to the mass loss. There are
indications that high mass-loss rates are only reached for relatively
young stars with massive progenitors.  The mass loss rates vary both
on long and short time scales: the short-term variations are likely
linked to luminosity variations associated with the thermal-pulse
cycle.  The  influence of mass loss in the post-AGB
phase is also discussed.
\end{abstract}

\section{The Asymptotic Giant Branch}

That Asymptotic Giant Branch (AGB) marks the end point of the
evolution of stars with initial masses $\lsim 8 \msolar$.  Stars on
the AGB are characterized by a degenerate C/O core, with nuclear
burning taking place in a shell; the total AGB life time is of order
$10^6\,$yr.  A recommended review of AGB evolution can be found in
Iben \&\ Renzini (1984), although many of the numerical results have
been updated since (e.g. Vassiliadis \&\ Wood 1993).

During the first 90\% of the total AGB life time the energy is
provided by helium shell burning.  This phase is called
the early AGB, and
is marked by a steady increase in luminosity.  The early AGB ends when
the helium, itself the remnant of earlier hydrogen burning, is finally
exhausted and hydrogen is re-ignited in a thin shell. This starts the
thermal-pulsing AGB: after sufficient helium has formed, the helium
ignites with a flash (the thermal pulse; TP) followed by a new phase
of helium burning. After the helium is again exhausted the cycle
reruns. Typical time scales between thermal pulses are from $10^4\,$yr
for the most massive stars to $>10^5\,$yr at low masses.

During the AGB the mass of the envelope surrounding the
nuclear-burning shell is steadily reduced, by nuclear burning on the
inside (increasing the mass of the C/O core) and mass loss through a
stellar wind on the outside.  When the remaining envelope mass has
dropped below $0.05\msolar$, the stellar temperature begins to
increase and the so-called post-AGB phase starts. The temperature can
reach values of $10^5\,$K (depending on mass) before nuclear burning
ceases and the star enters the white-dwarf cooling track.

The best-known AGB stars are the Miras and OH/IR stars. Miras have
typical luminosities slightly below $10^4\,\lsolar$. They are
large-amplitude pulsators with periods of 200--600 days. The periods
are generally stable, but the amplitudes may vary in time.  The
observed luminosities of Miras, shown in Table I, indicate that they
are located at the tip of the AGB. They are consistent with expected
values for the TP-AGB even for the faint globular cluster Miras
(Vassiliadis \&\ Wood 1993, their table 2).  The OH/IR stars are
optically obscured stars, with very high mass-loss rates and long
periods of 1000--3500 days. It has been a long-standing problem
whether they are more massive than the Miras or form a later
evolutionary phase (e.g. Habing 1989).

\begin{table}
\caption{Mira bolometric magnitudes}
\begin{center}
\begin{tabular}{lll}
\hline
population & $M_{bol}$ & ref. \\
\hline\\
Galactic Centre & $-4.5\,$--$\,-6$ & Jones et al. 1994 \\
Bulge           & $-3.5\,$--$\,-5$ & Whitelock et al. 1991 \\
Galactic Cap    & $-3.5\,$--$\,-5$ & Whitelock et al. 1994 \\
Metal-rich globular clusters & $-3\,$\phantom{.0}--$\,-4.5$ &
 Menzies \&\ Whitelock 1985 \\
LMC             & $\ge -7.2$ & Wood et al. 1992 \\
\hline\\
\end{tabular}
\end{center}
\end{table}

In this review I will first discuss the luminosity variations on the
AGB and the pulsations. The mass loss is discussed in Section 4,
followed by some comments on the initial--final mass relation and
the post-AGB evolution.

\section{Luminosity variations}

The luminosity during shell burning is in principle a monotonous
function of core mass. Core mass--luminosity relations have been
derived by many authors: a review which is still useful can be found
in Boothroyd \&\ Sackman (1988b).  From this one derives the classical
AGB limit of $M_{bol}=-7.2$ for the highest possible core mass of
$1.4\msolar$. This relation is calculated for quiescent hydrogen
burning.  Thus, it cannot be used for phases of helium burning, and
during the first thermal pulses the quiescent luminosity is in fact
not reached.  The core mass here refers to both the C/O core and the
helium shell.

Strong deviations from the $M_c L$ relation occur at times. First,
there is a strong effect on $L$ during the TP, as shown in Figure 1.
A substantial luminosity dip occurs during quiescent helium burning
which lasts up to 30\% of the cycle (Boothroyd \&\ Sackmann 1988a,
Vassiliadis \&\ Wood 1993).  For stars with low envelope mass
(i.e. all low-mass stars) the TP causes a brief but large luminosity
increase lasting $\lsim 10^3\,$yr. For stars with high envelope mass
this increase is much less pronounced. Second, for massive stars with
high-mass envelopes, the base of the convective envelope may penetrate
the hydrogen-burning layer.  This will temporarily increase the
luminosity of the star (Bl\"ocker \&\ Sch\"onberner 1991), although by
how much is not clear (Vassiliadis and Wood 1993).  When the envelope
mass decreases, deep convection ceases and the star will appear to
evolve {\it down} the AGB. Because of these deviations it is dangerous
to derive a core mass from observed AGB luminosities for individual
stars.

\begin{figure}
\vspace{11cm}
\caption{(Taken from Vassiliadis \&\ Wood 1993) Variations of
stellar parameters during the thermal-pulse cycle. The period is
calculated assuming fundamental mode; $V_{exp}$ is derived from the
period. $\dot M_6$ is the mass-loss rate in units of $10^{-6} M_\odot
yr^{-1}$  }
\end{figure}

Wood et al. (1992) argue that there is no evidence for overluminous
AGB stars in the LMC which might be expected from envelope
burning. This is based on the lack of stars with
$M_{bol}<-7.2$. However, AGB core masses possibly do not exceed
$1.1\msolar$, as indicated by initial--final mass relations
(e.g. Weideman 1987). If this is the case, the limiting magnitude
derived from the luminosity--core mass relation is closer to $-6.9$
and there would be some overluminous stars in the LMC.

\section{Pulsations and the $PL$ relation}

Optical Miras show a strong period--luminosity relation, in the sense
that higher luminosity coincide with longer periods. The relation is
quite narrow in the K-band; it shows more scatter when the bolometric
luminosity is used.  The relation has been interpreted as an
evolutionary sequence, but this is unlikely for several reasons: (1)
Mira life times ($\sim 10^5\,$yr) are too short to give an appreciable
evolution in $L$. (2) Miras in individual globular clusters show a
very small range in luminosity.  (3) For Galactic Miras the velocity
dispersion is a function of period (Table II, from Feast 1989)
indicating older progenitors for shorter-period Miras. (The
shortest-period Miras ($P<145\,$days) have lower velocity dispersion
than expected and may be a different group of objects, possibly
higher-overtone pulsators.) Thus, it appears that the relation traces
stars with different progenitor mass, with longer periods implying
younger, more massive stars. Metallicity may also play a role.

\begin{table}
\caption{(From Feast 1989) Mira velocity dispersions}
\begin{center}
\begin{tabular}{llll}
\hline
Period range & asymmetric drift & $\sigma_T$ & no. of stars\\
 (days)      &  (km/s)          & (km/s) \\
\hline\\
$<140$  &  $-33\pm13$ &  81 & 22 \\
145--200 & $-111\pm22$ & 180 & 46 \\
200--250 & $-61\pm22$  & 101 & 71 \\
250--300 & $-33\pm10$ &  88  & 77 \\
300--350 & $-32\pm6$  &  69  & 83  \\
350--410 & $-23\pm8$  &  58  & 54 \\
$>$410  &  $-15\pm8$  &  50  & 35 \\
\hline\\
\end{tabular}
\end{center}
\end{table}

Whitelock (1986) has shown that in globular clusters, semiregular
variables and Miras define a sequence in the $PL$ diagram which is
much shallower than the Mira $PL$ relation (Figure 2).  The sequence
has the same slope as found in recent theoretical calculations
(Vassiliadis \&\ Wood 1993, their figure 20) although it is shifted
towards shorter periods. It seems likely that the Whitelock relation
is the true evolutionary sequence and that the semi-regular variables
form a pre-Mira phase.

\begin{figure}
\vspace{7cm}
\caption{(Taken from Feast 1989) The Miras and SR variables in
globular clusters, together with the pre-Mira evolutionary track
suggested by Whitelock (1986). The Mira period--luminosity relation
(Feast et al. 1989) is shown for comparison.}
\end{figure}

During the thermal-pulse cycle, the changing characteristics of the
star will affect the periods. This is shown in Figure 1: the star will
move up and down (but not away from) the $PL$ relation. This behaviour
would of course not be observed if the Mira phase is confined to a
single phase of the thermal-pulse cycle.  A large shift up the $PL$
could occur during the luminosity spike following a TP. Even though
this spike is very brief, Mira pulsations could be triggered by a
resonance because the nuclear time scale of the pulse is of the same
order as the pulsation.  In conclusion, although the PL relation does
not appear to be an evolutionary sequence, movement up and down the
relation could still occur.

The $PL$ relation holds for $P<400\,$days. For longer periods the
scatter in $P$ becomes very large and in fact it is not obvious
whether any relation still exists (e.g. Wood et al. 1992, their figure
8). The relation could be broadened by envelope removal through mass
loss which would increase the period; this could be especially
important for more massive stars. A mode change to a lower pulsation
mode would also have this effect.  In contrast, contraction of the
star, which occurs at the start of the post-AGB evolution will shorten
its period.\\
\\
{\it Pulsation mode.} Whether Miras pulsate in the fundamental mode or
in an overtone has been a long-standing problem.  The existence of a
clear P-L relation implies that all Miras with $P<400$ days have the
same pulsation mode. (For the long-period OH/IR stars the absence of a
clear relation does not allow one to extrapolate this conclusion.) In
recent years the fundamental mode has generally been favoured: the
evidence in favour is reviewed by Wood (1989) and is mainly based on
observed shock velocities. However, the observed $T_{eff}$ of the
stars is lower than predicted from the fundamental mode. This was
recently confirmed by angular diameter measurements of R Leo (Tuthill
et al. 1994) which can exclude the fundamental mode for this
Mira. There are also a few Miras known with multiple periods. In the
case of IZ Peg, the periods are 488 and 345 days (Whitelock et
al. 1994): if these are both modes of the star, the fundamental mode
would have a period close to 1000 days.  Finally, from a Fourier
analyses of two Miras, Barth\'es \&\ Tuchman (1994) favour
first-overtone pulsations.  Thus, although the situation is by no
means clear, the balance is shifting to Miras being overtone
pulsators.

Whitelock (1986) has suggested that metal-poor stars pulsate in the
fundamental mode, while metal-rich stars pulsate in the first
overtone. Since only metal-rich clusters contain Miras, this would
imply that all Miras pulsate in the overtone.

\section{Mass loss}

Mass loss occurs along both the first giant branch (RGB) and the AGB,
but only reaches catastrophic values at the tip of the AGB. Here
values of $10^{-5} M_\odot yr^{-1}$ or higher are reached, which means
that the envelope is removed much faster by mass loss than by nuclear
burning. Thus, catastrophic mass loss will terminate the AGB,
and the time at which it occurs determines the final core mass.

The IRAS colours of AGB and RGB stars show a gap between stars with
essentially stellar colours at 12, 25 and 60$\mu$m, and 'dusty' stars
with significant excess (van der Veen \&\ Habing 1988). Fitting the
infrared spectra with a dust model, it appears that the gap coincides
with mass-loss rates $\lsim 10^{-7} M_\odot yr^{-1}$.
In contrast, optical
and UV observations show that essentially all red-giant stars show
mass loss, often with mass-loss rates as low as $10^{-10} M_\odot
yr^{-1}$ (e.g.  Judge \&\ Stencel 1991). These low mass-loss rates are
not seen in the infrared and appear to be 'dustless'.

All Miras have 'dusty' IRAS colours and thus exhibit substantial
mass-loss rates. Semi-regular variables, on the other hand, are found
in both categories. If we identify these with pre-Mira stars, the
conclusion is that 'dusty' mass loss begins during the semi-regular
phase and that the mass-loss rate correlates with the amplitude of the
pulsation. Very high mass-loss rates are only found for long-period
Miras with $P>400$ days, as shown by Jura (1986) for carbon stars and
Whitelock et al. (1994) for oxygen stars. Combining this with the
results from the previous section, this implies that high mass-loss
rates are in general not reached for the lowest-mass stars; for
instance, it is doubtful whether our Sun will ever become such a
long-period Mira.  However, low-mass stars could mimic this behaviour
during the brief high-luminosity phase following a thermal pulse.

At present there is no theoretical model which can actually predict
the mass loss. However, it is generally supposed that mass loss occurs
in two stages. One mechanism such as pulsations or other long-term
variability heats the atmosphere and causes it to extend to several
stellar radii. At that distance dust forms and the radiation pressure
on the dust will accelerate the shell and amplify the mass-loss
rate. This model naturally leads to the expectation that there are two
evolutionary phases: (1) At low mass-loss rates, the densities are so
low that even if dust still forms,
it will not be dynamically coupled to the
gas. In this case mass loss is driven solely by the mechanism which
heats the envelope; (2) At higher mass-loss rates the gas couples to
the gas and mass loss becomes more efficient and larger. The first
phase would then be associated with the 'dustless' RGB and early AGB,
the second phase is when the mass loss suddenly becomes
catastrophic. Calculations on the dust-to-gas coupling have been done
by MacGregor \&\ Stencel (1992) and Netzer \&\ Elitzur (1993).  In
either case, in principal the mass-loss rate would be determined by
the fundamental stellar parameters, and could be quantified.

Such parametrizations are indeed available.
The oldest is the well-known Reimers law (1975):

$$
\dot M = -4\times 10^{-13} {{LR}\over{M}}\quad M_\odot yr^{-1}
$$

\noindent with $L$,$M$,$R$ in solar units.
This relation was derived from data
on K stars and is known to predict too low rates from late AGB
stars. Fitting model results from Bowen (1988), Bl\"ocker (1993) has
proposed a stronger dependence on $L$ and $M$:

$$
\dot M = -4.8 \times 10^{-9} {{L^{2.7}}\over{M^{2.1}}}
{{LR}\over{M}}\quad M_\odot yr^{-1}
$$

\noindent This, however, predicts mass-loss rates
close to $10^{-3} M_\odot
yr^{-1}$ which are not observed. Although it is possible that the
phase with such high mass-loss rates lasts too short to be observable,
from stellar population studies in the LMC, Groenewegen and de Jong
(1994b) and Groenewegen et al. (1994) conclude that this formula
overestimates mass-loss rates by factors of 3--10.

The best observationally-established relation is given by Judge \&\
Stencel (1991) for a sample of RGB and AGB stars. They find a good
correlation between $\dot M$ and $\log g$, which can be converted to a
similar relation as above:

$$ \dot M = (2.3\pm 1.3)\times 10^{-14} \left( {{R^2}\over{M}}
\right)^{1.43\pm0.23}\quad M_\odot yr^{-1} $$

\noindent This is close to Reimers' law with
$L$ replaced by $R$, a reasonable
substitution since the two appear to be correlated during this phase
of the evolution. Although the observational spread in the relation is
significant, and especially the luminous carbon stars at the tip of
the AGB fall above the relation, it is probably the best one to use in
model calculations.

Vassiliadis \&\ Wood (1993) have suggested that momentum in the wind
should not exceed that in the radiation field, which for AGB stars
implies $\dot M \lsim 3 \times 10^{-5} M_\odot yr^{-1} $. They
calculate mass-loss rates from a $PL$ relation which is
very steep, and Groenewegen \&\ de Jong (1994b) find that their
results do not fit the observed stellar population in the LMC.
However, it is possible that the momentum limit should be included in
the above relations.\\
\\
{\it Mass-loss variations.} The above formulae predict that (1) $\dot
M$ increases with time on the AGB, (2) significant variations should
occur following a thermal pulse due to the change in surface
parameters. Thus, both long-term and short-term variations are
expected. The short-term variations occur on time scales comparable to
or shorter than the expansion of the nebula and should be observable.

The luminous OH/IR stars, which show very high mass-loss rates, often
show very faint CO emission (Heske et al. 1990).
Schutte \&\ Tielens (1989)
interpret this as evidence that the outer shell, where the CO emission
originates, traces a lower mass-loss phase. The increase in $\dot M$
would date back to $\sim 10^4$yr ago. These time scales are consistent
with those of the thermal pulse cycle, although it is not proven that
the increase is related to this.

Figure 3 shows the IRAS colours for all M stars from the GCVS, using
only data with good quality at all plotted bands ($q=3$). The vertical
lines connect points with the same mass-loss rates; horizontal lines
connect points with the same outer radius, converted to a time using
an expansion velocity of $10\,$km/s.
There are at most very few sources
for which the mass loss has started less than $10^3\,$yr ago. At
100$\mu$m almost all sources show a significant excess: this is either
explained by a very shallow emissivity index
(Rowan-Robinson et al. 1986)
or an outer shell  caused by earlier mass loss (e.g. van der Veen et
al. 1994). Thus, mass loss generally continues for $10^3$yr or longer.

\begin{figure}
\vspace{19cm}
\caption{IRAS colour-colour plots for M-stars taken from the GCVS.
These will be mainly (but not exclusively) AGB and RGB stars. Model
tracks at constant $\dot M$ (vertical lines) and constant time scale
(horizontal lines) are shown. (The absolute values for  $\dot M$
can change with assumed gas-to-dust ratio.)}
\end{figure}

Interruptions of mass loss appear to be a common feature
of optical carbon stars, which all show  60$\mu$m excess
indicating a detached, cool shell.  Willems \&\ de Jong (1986) have
proposed that this occurs when due to a thermal pulse the C/O ratio
becomes unity, with the resulting chemistry leading to a decrease in
dust formation. This scenario predicts that the detached shells are
oxygen rich. However, Bujarrabal \&\ Cernicharo (1994) and Groenewegen
\&\ de Jong (1994a) find evidence that at least for some carbon stars
the shells are carbon rich. Zijlstra et al. (1992) and Hashimoto
(1994) show that some oxygen stars also show detached shells. This
suggests that mass-loss interruptions may be a common phenomenon on
the AGB and are not limited to carbon stars. It would seem logical to
associate these interruptions with the phase of quiescent helium
burning during the TP cycle, when the luminosity is lowest (e.g.
Olofsson et al. 1990).  Zijlstra et al. conclude that mass loss on the
AGB contains a {\it periodic} component.\\
\\
{\it Implications} In conclusion, high mass-loss rates coincide with
high luminosity. This indicates that the primary factor determining
the mass-loss rate is stellar mass, although low-mass stars could
briefly obtain these luminosities during a thermal pulse.  If mass los
varies through the TP cycle, as is likely, stars may pass through
several mass-loss episodes, each lasting of order $10^4$yr.  Because
of the dust/gas decoupling, there may be a limit below which $\dot M$
may drop by one or more orders of magnitude. This could amplify the
effect of the TP-cycle.

\section{Initial--final mass relation}

The mass loss determines the mass of the core of the final white
dwarf, and thus the initial--final mass relation. This relation is
poorly constrained by observations: the only limits are given by a few
white dwarfs in young clusters (Weideman 1987, Vassiliadis \&\ Wood
1993). Applying the mass-loss relations above to theoretical models of
AGB stars (Groenewegen\&\ de Jong 1994b) leads to results consistent
with these limits.  Specifically, the relation is flat for
1--3$M_\odot$ stars which all end up as white dwarfs with masses
around 0.6$M_\odot$. Higher-mass stars end up as heavier white dwarfs,
but the Chandresekhar limit is never reached: the most massive AGB
stars leave 1--1.2$M_\odot$ remnants. This has important consequences
for the white-dwarf mass distribution. Han et al. (1994), following
earlier suggestions, have shown that the observational constraints are
well matched if one assumes that the envelope is lost shortly after
the binding energy of the envelope becomes positive. Their relation
could possibly be taken as a lower limit to the initial-final mass
relation.

\section {Post-AGB}

The AGB ends when the stellar temperature begins to rise: the
following phase is normally called the post-AGB. The mass loss
continues into the early post-AGB phase as indicated by observational
time scales: if the decrease in envelope mass (which determines how
fast the stellar temperature increases) were only due to nuclear
burning, it would take so long to reach the planetary-nebula phase
that the nebula would long have disappeared. Sch\"onberner (1983) has
assumed that the high mass loss ends when the star first reaches
$T_{eff}=5500\,$K. This is consistent with the fact that there are
many optically visible post-AGB stars of type F and later, but almost
none of K (Waters et al. 1989). However, van der Veen et al. (1994)
present a few cases where a significant cooler star has a detached
shell. The details of how the mass loss ends are still very unclear.

In a close binary system where one of the components passed through
the AGB, it is possible that part of the shell may be retained in a
circum-binary disk. This mechanism has been invoked to explain the
extreme metal deficiency of some post-AGB stars (Van Winckel et al.
1992, Waters et al. 1992). It could also explain why
highly obscured post-AGB stars invariably have  bipolar shells: only
objects where part of the shell remains close to the star will show
high obscuration during the post-AGB phase (e.g. Siebenmorgen et al.
1994), and the binary nature of these objects could cause the
non-sphericity of the shells.

Because the helium-burning phase in the TP cycle is relatively short,
it is commonly assumed that almost all post-AGB stars have left the
AGB as hydrogen burners. However, Vassiliadis (1993) argues that
25--50\% of all post-AGB stars are helium burners, based on his
evolutionary models. The reason is that for high-mass stars, the time
the post-AGB star is visible is much longer for helium burners than
for hydrogen burners, thus biasing the statistics. In addition, for
low-mass stars the mass loss may strongly peak immediately after a
pulse thus increasing the likelihood the envelope will be lost at this
time. Therefore, both the highest and lowest-mass stars may often be
helium burners. The expected relative numbers of helium burners is
strongly depended on poorly understood details of
the mass-loss process.

For hydrogen-burning stars, there is a chance that a final thermal
pulse will occur during the post-AGB evolution or even in the
white-dwarf phase. It is thought that this may lead to a final phase
of high mass loss, in which the entire remaining hydrogen layer is
removed.  The mass loss in this phase will be hydrogen-poor leading to
strong infrared (dust) emission. Although a rare phenomenon, there are
in fact stars which appear to have very recently entered this phase,
in particular FG SGe (e.g. van Genderen 1994) and N66 in the LMC
(Pe\~na et al. 1994). Especially the latter is a good candidate for
studying this final episode of mass loss.

\section{References}
%\noindent {\bf References}
\begin{list}
{}{\itemsep 0pt \parsep 0pt \leftmargin 3em \itemindent -3em}

\item Barth\`es, D., Tuchman, Y., 1994, A\&A, 289, 429

\item Bl\"ocker, T., Sch\"onberner, D., 1991, A\&A, 244, L43

\item Bl\"ocker, T., 1993, Acta Astronomica, 43, 305

\item Boothroyd, A.I., Sackmann, I.-J., 1988a, ApJ, 328, 632

\item Boothroyd, A.I., Sackmann, I.-J.,  1988b,  ApJ,  328,  641

\item Bowen, G.H., 1988, ApJ, 329, 299

\item Bujarrabal, V., Cernicharo, J., 1994, A\&A, in press

\item Feast, M.W., 1989, in: The Use of Pulsating Stars in Fundamental
Problems of Astronomy, ed. E.G. Schmidt
(Cambridge University Press), p.205

\item Feast, M.W., Glass, I.S., Whitelock, P.A., Catchpole, R.M., 1989,
MNRAS, 241, 375

\item Groenewegen, M.A.T., de Jong, T., 1994a, A\&A, 282, 115

\item Groenewegen, M.A.T., de Jong, T., 1994b, A\&A, 283, 463

\item Groenewegen, M.A.T., van den Hoek, L.B., de Jong, T., 194,
A\&A, in press

\item Habing, H.J., 1989, in: From Miras to Planetary Nebulae:
which path for
stellar evolution? (Editions Fronti\`eres, Gif sur Yvette),  p.16

\item Han, Z., Podsiadlowski, Ph., Eggleton, P.P., 1994,
MNRAS, 270, 121

\item Hashimoto, O., 1994, A\&ASS, in press

\item Heske, A., Forveille, T., Omont, A., van der Veen, W.E.C.J.,
Habing, H.J., 1990, A\&A, 239, 173

\item Iben Jr., I, Renzini, A,  1983,  ARA\&A,  21,  271

\item Jones, T.J., Gehrz, R.D., Lawrence, G.F., McGregor, P.J., 1994,
University of Minnesota preprint

\item Judge, P.G., Stencel, R.E., 1991, ApJ, 371, 357

\item Jura, M., 1986, ApJ, 303, 327

\item MacGregor, K.B., Stencel, R.E., 1992, ApJ, 397, 644

\item Menzies, J.W., Whitelock, P.A., 1985, MNRAS, 212, 783

\item Netzer, N., Elitzur, M., 1993, ApJ, 410, 701

\item Olofsson, H., Carlstr\"om, U., Eriksson, K., Gustafsson, B.,
Willson, L.A., 1990, A\&A, 230, L13

\item Pe\~na, M., Torres-Peimbert, S., Peimbert, M., 1994,
ApJ, 428, L9

\item Reimers, D., 1975, in: Problems in Stellar Atmospheres and
Envelopes, Basheck et al., eds. (Springer, Berlin) p.229

\item Rowan-Robinson, M., Lock, T.D., Walker, D.W., Harris, S.,
1986, MNRAS,
222, 273

\item Sch\"onberner, D., 1983, ApJ, 272, 708

\item Schutte, W.A., Tielens, A.G.G.M., 1989, ApJ, 343, 369

\item Siebenmorgen, R., Zijlstra, A.A., Kr\"ugel, E., 1994,
MNRAS, in press

\item Tuthill, P.G., Haniff, C.A., Baldwin, J.E., Feast, M.W., 1994,
MNRAS, 266, 745

\item van der Veen, W.E.C.J., Habing, H.J., 1988, A\&A, 194, 125

\item van der Veen, W.E.C.J., Waters, L.B.F.M., Trams, N.R.,
Matthews, H.E., 1994, A\&A, 285, 551

\item van Genderen, A.M., A\&A, 284, 465

\item Van Winckel, H., Mathis, J.S., Waelkens, C., 1992,
Nature, 356, 500

\item Vassiliadis, E., 1993, Acta Astronomica, 43, 315

\item Vassiliadis, E., Wood, P.R.,  1993,  ApJ,  413,  641

\item Weidemann, V.,  1987,  A\&A,  188,  74

\item Whitelock, P.A., 1986, MNRAS, 219, 525

\item Whitelock, P.A., Feast, M.W., Catchpole, R.M., 1991,
MNRAS, 248, 276

\item Whitelock, P.A. et al. 1994, MNRAS, 267, 711

\item Waters, L.B.F.M., Waelkens, C., Trams, N.R., 1989, in:
 From Miras to  Planetary Nebulae: which path for stellar evolution?
(Editions  Fronti\`eres, Gif sur Yvette), p.449

\item Waters, L.B.F.M., Trams, N.R., Waelkens, C., 1992, A\&A, 262, L37

\item Willems, F.J., de Jong, T., 1986, ApJ, 303, L39

\item Wood, P.R., 1989, in: From Miras to Planetary Nebulae:
which path for stellar evolution? (Editions Fronti\`eres,
Gif sur Yvette), p. 67

\item Wood, P.R., Whiteoak, J.B., Hughes, S.M.G., Bessell, M.S.,
Gardner, F.F., Hyland, A.R., 1992, ApJ, 397, 552

\item Zijlstra , A.A., Loup, C., Waters, L.B.F.M., de Jong, T.,
1992, A\&A, 265, L5

\end{list}

\end{document}